\documentclass[namedreferences]{SolarPhysics}
\usepackage{epsfig}

\newcommand{\be}{\begin{equation}}
\newcommand{\ee}{\end{equation}}
\newcommand{\beq}{\begin{eqnarray}}
\newcommand{\eeq}{\end{eqnarray}}

\newcommand{\rsun}{\mbox{$R_\odot$}}
\newcommand{\rad}{\mbox{rad\,m$^{-2}$}}

\begin{document}
\begin{article}
\begin{opening}

\title{The Magnetic Field of the Solar Corona from\\
Pulsar Observations}

\author{S. M. \surname{ORD}$^{1}$,
        S.   \surname{JOHNSTON}$^{2}$,
        J.  \surname{SARKISSIAN}$^{3}$
        }

\runningauthor{ORD, JOHNSTON AND SARKISSIAN}
\runningtitle{The magnetic field of the solar corona}

\institute{$^{1}$ School of Physics, University of Sydney, Sydney, NSW 2006, Australia
                  \email{steve.ord@gmail.com}\\
           $^{2}$ Australia Telescope National Facility, Epping NSW 1710, Australia\\
           $^{3}$ Australia Telescope National Facility, Epping NSW 1710, Australia\\
           }

\date{Received ; accepted }

\begin{abstract}
We present a novel experiment with the capacity to independently measure 
both the electron density and the magnetic field of the solar corona.
We achieve this through measurement of the excess
Faraday rotation due to propagation of the polarised emission from a
number of pulsars through the magnetic field of the solar corona.
This method yields independent
measures of the integrated electron density, via dispersion of the pulsed
signal and the magnetic field, via the amount of Faraday rotation.
In principle this allows
the determination  of the integrated magnetic field through the solar corona
along many lines of sight without any assumptions regarding the electron
density distribution.
We present a detection of an increase in the rotation measure 
of the pulsar J1801$-$2304 of approximately 160~\rad at an elongation 
of 0.95$^\circ$ from the centre of the solar disk. This corresponds to a 
lower limit of the magnetic field strength along this line of 
sight of  $> 393\mu\mathrm{G}$. The lack of precision in the integrated 
electron density measurement restricts this result to a limit, but 
application of coronal plasma models can further constrain this to 
approximately 20mG, along a path passing 2.5 solar radii from the solar limb. Which is consistent with predictions obtained using extensions to the Source Surface models published by Wilcox Solar Observatory

\end{abstract}
\keywords{Sun: magnetic field}

\end{opening}

\section{Introduction}
\label{Introduction} 
Our understanding of the solar corona is far from complete.  Attempts are
continually being made to marry theory and observations, but these efforts are
hampered by the difficulty in obtaining direct measurements of physical
properties throughout the corona. Without these measurements the state of the
corona must be inferred from models that extrapolate from those observations
that can be obtained from the solar surface or at distances
comparable to 1 AU. Experiments have attempted to probe the outer corona
via Faraday rotation through a variety of
methods including measuring the rotation of the position angle of polarised
emissions from the Cassini space probe (Jensen et al 2005)\nocite{jbi+05} and Pioneer 6
(Levy et al 1969)\nocite{lss+69}. Background radio sources have also been used to constrain the
mean integrated line of sight magnetic field (Mancuso \& Spangler 2000)\nocite{ms00b}. We have used the co-axial 10/50cm receiver attached to the 64--m Parkes radio telescope to observe a number of pulsars as the background radio
sources. Pulsars have a number of advantages over the background sources
observed successfully by Mancuso \& Spangler (2005), pulsars are polarised point sources
and the position angle (PA) of any linearly polarised component is routinely
measured by pulsar observers, their radiation is also broadband and pulsed,
allowing simultaneous observation at multiple frequencies and the direct
determination of the electron density along the line of sight. 

This method has been motivated by the success of the Cassini and background radio
source experiments and can in principle present a unique
opportunity to disentangle the two properties of the solar corona that
contribute to any observed rotation measure (RM). Namely that by measuring the
dispersion of the pulsar signal we can determine the coronal electron density
independently, this allows the path integrated magnetic field to be uniquely
determined without recourse to a theoretical model of the coronal density
profile. Consequently we can determine whether any variations in observed RM
are due to fluctuations in the field, or variations in electron density. This
has not been possible in the previous experiments of this nature. A further
motivation is the large number of simultaneous lines of sight 
through the corona that
can be observed within a given session, therefore the 3-dimensional
structure of the corona can be constrained to a considerable degree.

We carried out this experiment in 2006 December. We detected 
a large variation in rotation measure for one target, and smaller variations
in the second target.  No variation was detected in our calibrator 
pulsar, demonstrating the feasibility of the experiment and convincing
us of the veracity of our results.

This paper is structured in the following manner, we firstly present the level
of the effect we intend to measure, we determine this via a simple model
magnetic field constructed from a potential field model using data from the
Wilcox Solar Observatory and a model of the coronal electron density.  We then
discuss the details of the experimental method. We finally present the results of the experiment, together with the calibration observations

\section{Experimental Justification}
\label{sec:MODELS}

In order to explore the feasibility of this experiment we have constructed a
relatively simple model solar corona, using a potential field model and
separate electron density model. This construction was then used to examine the
level of the expected variations in RM and DM.
 
\subsection{The Potential Field Model}

Many approaches to model the solar magnetic field have been taken; the earliest
being the potential field model developed independently by Schatten~
et~al.~(1969) and Altschuler~\&~Newkirk~(1969)\nocite{swn69,an69}.  Which
extends photospheric measurements from magnetograms out to a {\it source
surface}, typically 2.5 R$_\odot$ from the centre of the sun. It is generally
not considered wise to simply extrapolate from the field values at this surface
out to arbitrary radii, as we have done here.  This however, is not intended to
be a perfect model of the field in the outer corona, but a general model that
maintains the geometry of the field and provides approximate values for the
field in a ``typical'' quiet sun.

Following Altschuler~\&~Newkirk~(1969), the potential field model begins with
the assumption that the coronal field is supporting no currents is equivalent
to: 

\begin{equation}
\nabla \times {\mathbf B} = 0.
\end{equation}
The magnetic field can therefore be represented as the negative gradient of the scalar potential, $\psi$,
\begin{equation}
{\mathbf B} = -\nabla\psi
\end{equation}
Since
\begin{equation}
\nabla \cdot {\mathbf B} = 0,
\end{equation} 
then
\begin{equation}
\nabla \cdot\nabla\psi = 0,
 \end{equation}
 which is equivalent to:
 \begin{equation}
 \nabla^2\psi=0.
 \end{equation}
 Therefore the potential satisfies Laplace's equation. A solution of Laplace's
equation can be found in terms of Legendre polynomials. In this case we have
used a series of Schmidt quasi-normalised Legendre polynomials. The  potential
field solution decomposed into spherical polar components, $r$, $\theta$,
$\phi$, where $\theta$ is the co-latitude, is of the following form:
\begin{eqnarray}
B_{r} & = & \sum_{lm}P^{m}_{l}(\cos{\theta})\left(g_{lm}\cos{m\theta} + h_{lm}\sin{m\theta}\right) \nonumber  \\
           & {} & \times  \left((l + 1)\left(\frac{\rsun}{r}\right)^{l+2} - l\left(\frac{r}{R_{s}}\right)^{l-1}c_{l}\right) 
\end{eqnarray}
\begin{eqnarray}
B_{\theta}  & = & -1 \times \sum_{lm}\left(\left(\frac{\rsun}{r}\right)^{l+2} + \left(\frac{r}{R_{s}}\right)^{l-1}c_{l}\right) \nonumber  \\  
         & & \times \left(g_{lm}\cos{m\theta} + h_{lm}\sin{m\theta}\right)\frac{\delta P^{m}_{l}(\cos{\theta})}{\delta \theta} 
\end{eqnarray}
\begin{eqnarray}         
B_{\phi}  & = & \sum_{lm}\left(\left(\frac{\rsun}{r}\right)^{l+2}  + \left(\frac{r}{R_{s}}\right)^{l-1}c_{l}\right)\frac{m}{\sin{\theta}} \nonumber  \\ 
& & \times (\cos{\theta})\left(g_{lm}\sin{m\theta} - h_{lm}\cos{m\theta}\right) P^{m}_{l}(\cos{\theta})
\end{eqnarray}
where:
\begin{equation}
c_{l} = -\frac{\rsun}{R_s}, 
\end{equation}
$R_s$ is the distance to the source surface (2.5 $\rsun$ in this case) and
$\rsun$ is the solar radius, the coefficients $h_{lm}$ and $g_{lm}$ are
tabulated by Wilcox Solar Observatory for the rotation corresponding to the epoch of our experiment.


\begin{figure}
\includegraphics[width=\linewidth]{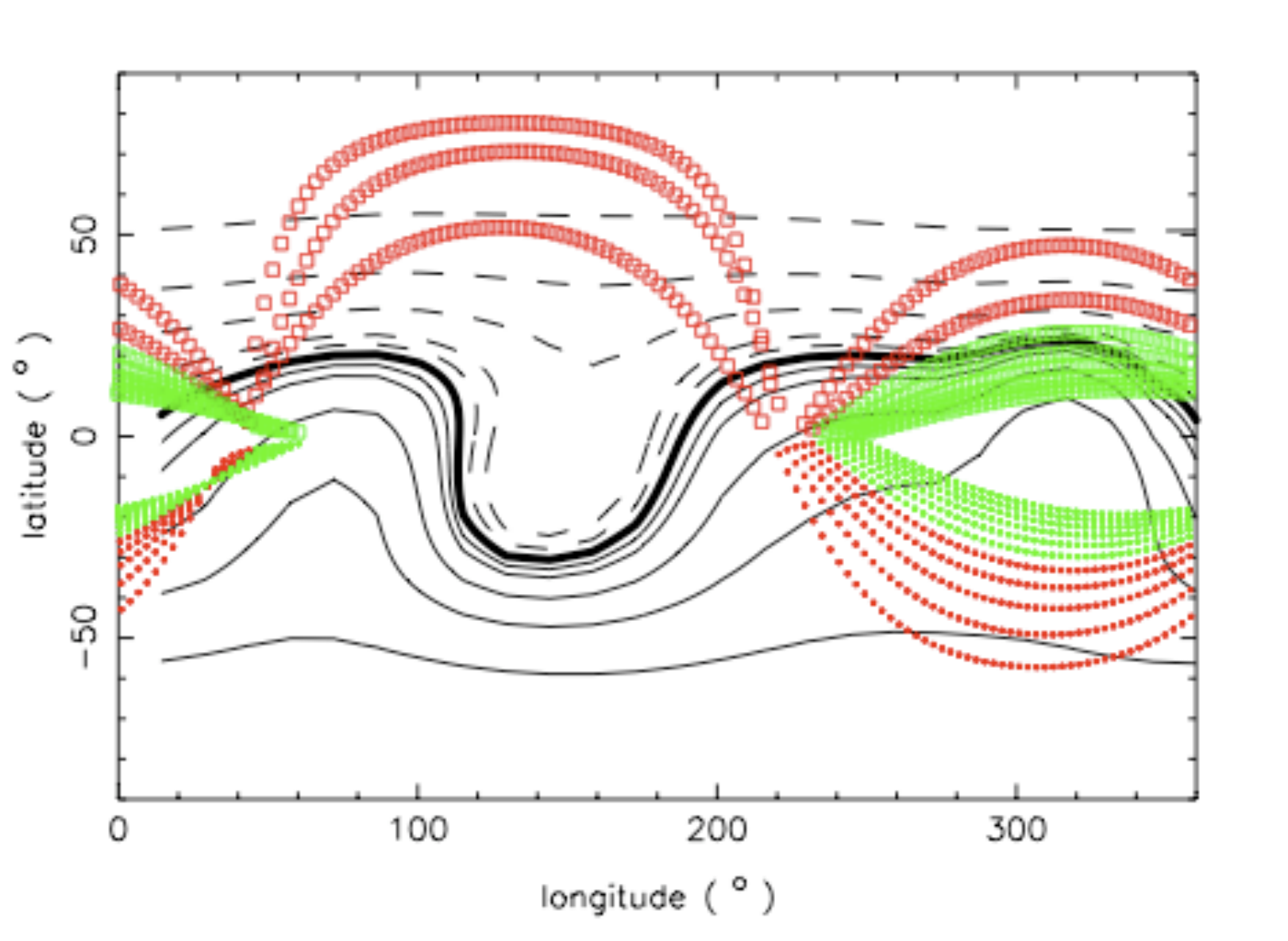}

\caption{The radial component of the magnetic field of the sun at a distance of 2.5
solar radii from the centre of the sun. Calculated using the sum of
Quasi-Schmidt Normalised Legendre polynomials. Using coefficients supplied by Wilcox Solar Observatory for Carrington rotation 2051. The contour lines are 0, $\pm$1, 2, 5, 10, 20 $\mu$T, the solid black line being the neutral line. The boxes represent the lines of sight to J1801-2304, the dots are lines-of-sight to J1757-2421. The red symbols denote lines of sight between the 22nd and the 23rd of December, and the green symbols the 23rd and 24th. }
\label{fig:source}
\end{figure}

\subsection{An Electron Density Model}
\label{electron_density}
We have applied the same electron model used by 
Mancuso \& Spangler (2000)\nocite{ms00b}. This is a two component model introduced by Guhathakurta et al. (1996)\nocite{ghm96}.
The two components are a coronal hole and an equatorial streamer belt. The
equatorial streamer confined to the region of zero
magnetic field (the solid line in Figure \ref {fig:source}).

The two components in the Mancuso and Spangler~(2000)\nocite{ms00b} version of
the Guhathukurta et al. (1996)\nocite{ghm96} model are designated $n_{CS}$ and $n_{CH}$, the {\it current
sheet} and the {\it coronal hole}. They are combined in the following manner to
obtain the total electron density, $w$ represents the width of the streamer  and $\theta$ the angular distance in latitude above or below it:
\begin{eqnarray}
\nonumber n (r,\theta,\phi) &=& n_{CH}(r) + [n_{CS}(r) - n_{CH}(r)] \\
                                 & & \times\, \mathrm{exp}[-\theta^{2} / w^{2}(r,\phi)].
\end{eqnarray}
Where,
\begin{equation}
n_{CH}(r) =  \left[16.15\left(\frac{r}{\rsun}\right)^{-4.39} + 9.975\left(\frac{r}{\rsun}\right)^{-4.09}\right.  \\  
          \left. + 1.099\left(\frac{r}{\rsun}\right)^{-2}\right] \times 10^{5}. 
\end{equation}
and the streamer ($n_{CS}$)  is given by:
\begin{equation}
n_{CS}(r) = \left[365\left(\frac{r}{\rsun}\right)^{-4.31} + 3.6\left(\frac{r}{\rsun}\right)^{-2}\right] \times 10^{5},
\end{equation}
the electron density being measured in cm$^{-3}$. We have made the assumption in this model that the streamer is $\pi / 16$ rad in width.

\subsection{The Pulsars} 

The ecliptic plane is tilted with respect to the
Galaxy, therefore at two times of the year the Sun passes through the Galactic
plane as it is projected on the celestial sphere. The known pulsar population
is preferentially distributed within the Galactic plane and interrogation of
the ATNF pulsar database (Manchester et al. 2005)\nocite{mhth05} has provided a list of pulsars that
pass close to the Sun (within 2.5 degrees) are grouped in 
ecliptic longitude and display appreciable flux in the frequency bands of interest. These pulsars are
listed in Table \ref{tab:PSR}. We only attempted observations of a small subset of these pulsars, but they are all included in the table for completeness.

\begin{table}
\caption{Possible Targets}
\label{tab:PSR}
\begin{tabular}{lll}
Pulsar & Ecliptic & Flux (mJy) \\
Name & Longitude : Latitude & at 1400~MHz \\
\hline \\
J1652-2404        &  254.73 : -1.50            & 1.1$^{a}$ \\
J1721-2457        &  261.18 : -1.81            & unknown\\
J1730-2304        &  263.19 : 0.19             & 4$^{b}$\\
J1733-2228        &  263.86 : 0.821            & 2.3$^{d}$\\
J1753-2501        &  268.52 : -1.576           & 2.3$^{d}$\\
J1756-2251        &  269.26 : 0.57             & 0.6$^{e}$\\
J1756-2435        &  269.30 : -1.1554          & 2.0$^{d}$\\
J1757-2421        &  269.42 : -0.9307          & 3.9$^{d}$\\
J1757-2223        &  269.50 : 1.04             & 1.1$^{c}$\\
J1759-2205        &  269.86 : 1.3467           & 1.3$^{d}$\\
J1759-2302        &  269.96 : 0.40             & 1.3$^{c}$\\
J1801-2304        &  270.31 : 0.36             & 2.2$^{d}$\\
J1803-2137        &  270.89 : 1.8282           & 7.6$^{d}$\\
J1807-2459A       &  271.66 : -1.57            & 1.1$^{d}$\\
J1817-2311        &  273.91 : 0.20             & unknown\\
J1822-2256        &  275.28 : 0.392            & 2.4$^{d}$\\
J2048-1616        &  310.12 : 1.49             & 13$^{a}$\\
\hline
\multicolumn{3}{c}{References:} \\
\multicolumn{2}{l}{a: Lorimer et al. (1995)\nocite{lylg95}} \\ 
\multicolumn{2}{l}{b: Kramer et al. (1998)\nocite{kxl+98}} \\ 
\multicolumn{2}{l}{c: Morris et al. (2002)\nocite{mhl+02}} \\
\multicolumn{2}{l}{d: Hobbs et al. (2004)\nocite{hfs+04}} \\
\multicolumn{2}{l}{e: Faulkner et al. (2005)\nocite{fkl+05}} \\
\end{tabular}
\end{table}

\subsection{Pulsar Observations}

Pulsars are point sources of periodic broad band noise. A pulsar observation
consists of an integration, lasting typically hundreds or thousands of pulse
periods, which is folded modulo the topocentric pulse period to obtain a pulse
profile. Pulsar emission is in general polarised,
but different pulsars display different degrees of circular and linear
polarisation. The position angle of linear polarisation varies through the
pulsar profile, but is stable in time.

Changes in the Faraday rotation that the radiation from pulsars is subjected to
during passage through the solar corona manifests itself as a rotation of the
polarisation angle of linear polarisation. This change in position angle is a
function of frequency and is apparent both across the band and between
observing bands. Changes in the electron content along the line of sight
manifests itself as a dispersive delay between the arrival of the pulse at
different frequencies.

\subsubsection{Measuring Dispersion and Rotation Measure}

In routine pulsar observations the ionised content of the interstellar medium (ISM)
results in a dispersive delay between pulse arrival times at different
frequencies. For this experiment there is a further delay due to the electron
content in the solar corona. Although this is a small effect it is measurable
if the same pulse can be observed over a considerable frequency interval.
Measurement of this delay provides an independent measure of the electron
content along the line of sight from the following relation:
\begin{equation}
\Delta t = \frac{e^{2}}{2 \pi m_{e}c} \times \left(\frac{1}{f_{1}^{2}} - \frac{1}{f_{2}^{2}} \right) \times \int_{0}^{d} n_{e} \mathrm{d}l,
\end{equation}
which is generally expressed as:
\begin{equation}
\Delta t \simeq 4.15 \times 10^{6} \mathrm{ms} \times \left(\frac{1}{f_{1}^{2}} - \frac{1}{f_{2}^{2}}\right) \times \mathrm{DM}
\label{EQ:DM}
\end{equation}
where both frequencies are in MHz, and the dispersion measure (DM) is expressed
in cm$^{-3}$pc.  

The rotation measure (RM) is defined as:
\begin{equation}
\mathrm{RM} = \frac{e^3}{2 \pi m^{2}_{e} c^{4}} \times \int_{0}^{d} n_{e}\mathbf{B}\cdot \mathrm{d}\mathbf{l}
\label{EQ:RM}
\end{equation}
The total Faraday rotation is integrated along a path, and the total angular
rotation in polarisation position angle being given by:
\begin{equation}
\Delta \Phi = \lambda^{2} \times \mathrm{RM},
\end{equation}
It is readily appreciated that if both RM and DM can be measured then the
integrated magnetic field strength along the path can be obtained via:
\begin{equation}
\langle \mathrm{B}_{\|} \rangle = 1.23 \mu \mathrm{G} \left(\frac{RM}{\mathrm{rad}\,\mathrm{m^{-2}}} \right) \left( \frac{DM}{\mathrm{cm}^{-3}\mathrm{pc}} \right)^{-1}
\label{B}
\end{equation}

\section{The Experiment}
\subsection{Observations}
The observations were carried out using the Parkes radio telescope
between 2006 December 20 and 28. Of the pulsars listed in Table~1
we observed PSRs J1801$-$2304, 1757$-$2421, 1757$-$2223 and J1822$-$2256.
Of these, only the first two were useful for this experiment;
the latter two had very low linear polarization and it was very
difficult to measure an accurate RM. We also observed a
calibrator pulsar, PSR J1644$-$4559 whose properties are well
known (Johnston 2004\nocite{joh04}) and which is far enough away
from the Sun to undergo no changes in its parameters.

We observed at 3 different wavebands, 50cm, 20cm and 10cm, using
two different receiver packages. The first was the H-OH receiver
at a central frequency of 1369~MHz with a bandwidth of 256~MHz.
We also used the 10/50~cm receiver, a dual frequency system capable
of observing simultaneously at both 650 and 3000~MHz.
We used central frequencies of 3100~MHz with a bandwidth of 512~MHz
and 690~MHz with an effective bandwidth (after interference rejection)
of 35~MHz.

All receivers have orthogonal linear feeds and also have a pulsed
calibration signal which can be injected at a position angle of
45 degrees to the two feed probes.  A digital correlator
was used which subdivided the bandwidth into 1024 frequency channels
and provided all four Stokes' parameters. We also recorded 1024
phase bins per pulse period for each Stokes' parameter.
The pulsars were observed for $\sim$30 minutes on each occasion and
prior to the observation of
the pulsar a 3-min observation of the pulsed calibration signal was
made.  The data were written to disk in FITS format for subsequent
off-line analysis.

Data analysis was carried out using the PSRCHIVE
software package (Hotan et al. 2004)\nocite{hvm04} and the 
analyis and calibration were carried out in an identical fashion
in that described in detail in Johnston et al. (2005)\nocite{jhv+05}.  Most importantly,
we are able to determine absolute position angles (PA) for the linearly
polarized radiation at all three of our observing frequencies.

\subsection{Results}
\begin{table}
\caption{Results for PSR J1757$-$2421}
\begin{tabular}{lccl}
\hline
Date & Elongation & DM & RM \\
 & (deg) & (cm$^{-3}$pc) & (rad m$^{-2}$) \\
\hline
Dec 20 & 1.89 & 179.5 $\pm$0.12 & $-$33$\pm$2 \\
Dec 21$^{\mathrm{b}}$ & 1.10 & 177.4$\pm$0.5 & $-$26$\pm$2\\
Dec 22$^{\mathrm{a}}$ &    1.11  & $--$ & $--$ \\
Dec 23 & 1.96 & 179.9$\pm$0.2 & $-$26$\pm$2\\
Dec 24 & 2.88 & 179.8$\pm$0.2 & $-$26$\pm$2\\
Dec 28 &   8.43  & 179.4$\pm$0.06 & $-$26$\pm$2 \\
\hline
\\
\end{tabular}

a: Not observed\\
b: Very weak \\
\label{res1757}
\end{table}

\begin{table}
\caption{Results for PSR J1801$-$2304}
\begin{tabular}{lccl}
\hline
Date & Elongation & DM & RM \\
& (deg) & (cm$^{-3}$pc) & (rad m$^{-2}$) \\
\hline
Dec 20 & 2.60 & 1071.9$\pm$0.4 & $-$1170$\pm$7 \\
Dec 21$^{\mathrm{a}}$ & 1.50 & $--$ & $--$ \\
Dec 22$^{\mathrm{b}}$ & 0.44 & $--$ & $--$ \\
Dec 23 & 0.96 & 1070.1$\pm$0.5 & $-$977$\pm$7 \\
Dec 24 & 1.84 & 1069.2$\pm$0.4 & $-$1145$\pm$7 \\
Dec 25 & 2.91 & 1070.0$\pm0.3$ & $-$1160$\pm$7 \\
Dec 27 & 5.11 & 1070.2$\pm$0.4& $-$1158$\pm$7 \\
\hline
\\
\end{tabular}

a: Not observed\\
b: Very weak \\

\label{res1801}
\end{table}

The absolute polarization calibration at all three observing frequencies
allows us to obtain a more accurate RM in the following way.
First we obtain the RM through fitting to the PAs of
the linear radiation as a function of observing frequency across
the band at a single frequency.  We then improve the RMs by comparing
PAs between the different frequency bands, where the large lever
arm yields a much smaller error bar.
Unfortunately, this is not possible for PSR J1801$-$2304 because
the pulsar is essentially invisible at our lowest frequency. We
therefore rely on the 3.1~GHz data to deduce the RM.

The dispersion measure (DM) was computed by first measuring the time of
arrival of the pulsar as a function of frequency across the 
observing band. Then equation \ref{EQ:DM} was fitted in the arrival times to
determine an accurate value of DM.

Results for PSRs J1757$-$2421 and J1801$-$2304 are shown in
tables \ref{res1757} and \ref{res1801} respectively.
There is a surprising difference between the two cases. For
PSR~J1757$-$2421, the RM of the pulsar far from the Sun
is $-26$~rad\,m$^{-2}$ and we only see deviations from this value at
a single epoch (Dec 20). For PSR~J1801$-$2304 in contrast we 
see significant variations from day to day. The RM of the pulsar far
from the Sun is $-$1150~rad\,m$^{-2}$ and this value was also obtained
once the pulsar was more than 5 degrees from the Sun.
However, on Dec 23, when the pulsar was less than 1 degree from the
Sun's centre, we recorded an RM change of 160~rad\,m$^{-2}$.
On Dec 22, the date of the closest approach to the Sun, the total intensity
of the pulsar was low (likely caused by an increase in the system temperature
because of the proximity to the Sun) and we were unable to determine
an RM value.

This is an extremely high RM variation, although not 
unprecedented in the pulsar literature. PSR B1259$-$63 is in an
eccentric orbit about a Be star companion (Johnston et al. 1992)\nocite{jml+92}. During the
periastron approach, the RM changes significantly, reaching a peak
variation in excess of 10$^4$ rad\,m$^{-2}$ (Conners et al. 2002; Johnston et al. 2005)\nocite{cjmm02,jbwm05}. 
The implications of this change are that the magnetic field in the
vicinity of the Be star of $\sim$6~mG.

\subsection{The Predictions}

\begin{figure}
\includegraphics[width=\linewidth]{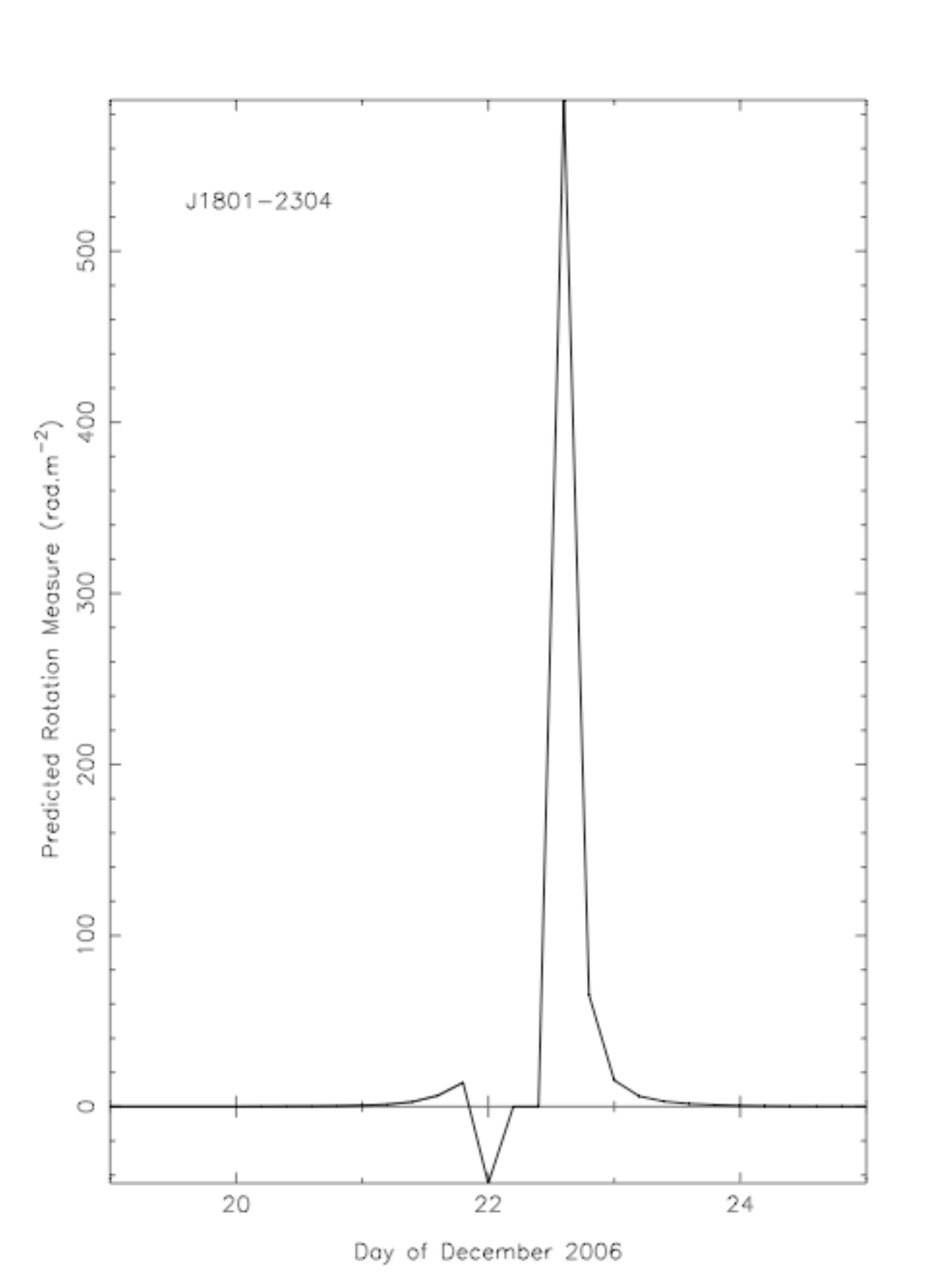}
\caption{The predicted excess Faraday rotation for J1801$-$2304. The pulsar was observed throughout this period, unfortunately the observation at the height of the predicted Faraday rotation has insufficient signal to noise ratio to allow a confirmation} 
\label{fig:PSRFARA}
\end{figure}

The simple models described in section \ref{sec:MODELS} have been used to
model this experiment. The line of sight to each pulsar as the
Sun moves across the sky has been projected through the model solar 
corona and the change in measured electron content and excess Faraday 
rotation predicted as a function of observing day. Figure \ref{fig:PSRFARA} 
presents the expected Faraday
rotation at the three observing wavelengths. The predictions 
indicate that a large change in rotation measure for J1801$-$2304 was 
actually expected for the days in question. However the observed RM 
change is higher even than the prediction suggesting that our simple 
model is not fully describing the state of the magnetosphere.  The 
predictions suggest that an even greater effect would have been 
observed around the 22nd of December, however as noted previously, 
observations taken on that day has a signal to noise ratio too low to 
detect a variation in RM.  


\begin{figure}
\includegraphics[width=\linewidth]{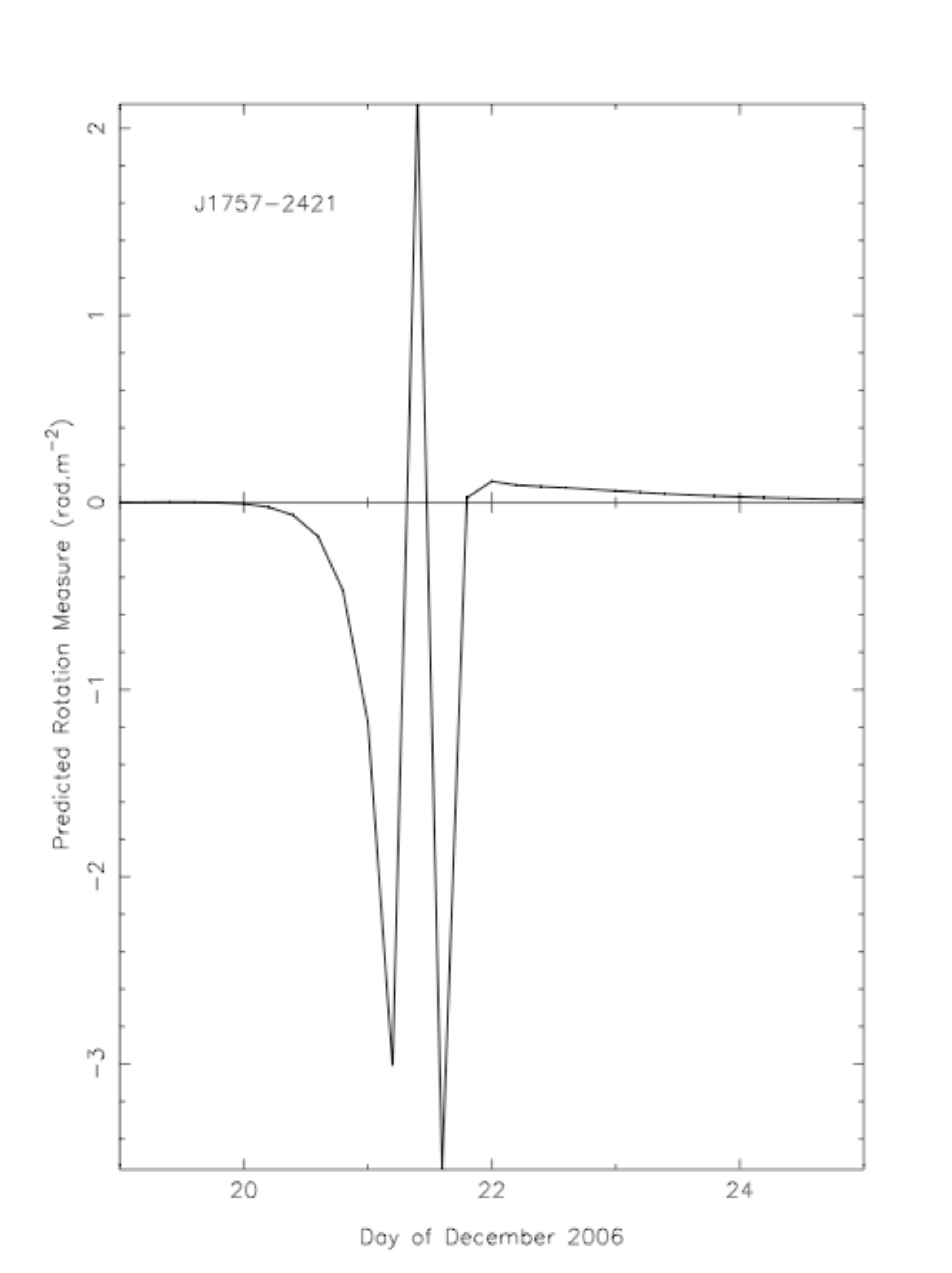}
\caption{The predicted excess Faraday rotation for J1757$-$2421. The line of sight to this pulsar traverses regions of the corona that are symmetric in their magnetic field and electron properties, therefore it displays little net Faraday rotation.} 
\label{fig:PSRFARA2}
\end{figure}

\begin{figure}
\includegraphics[width=\linewidth]{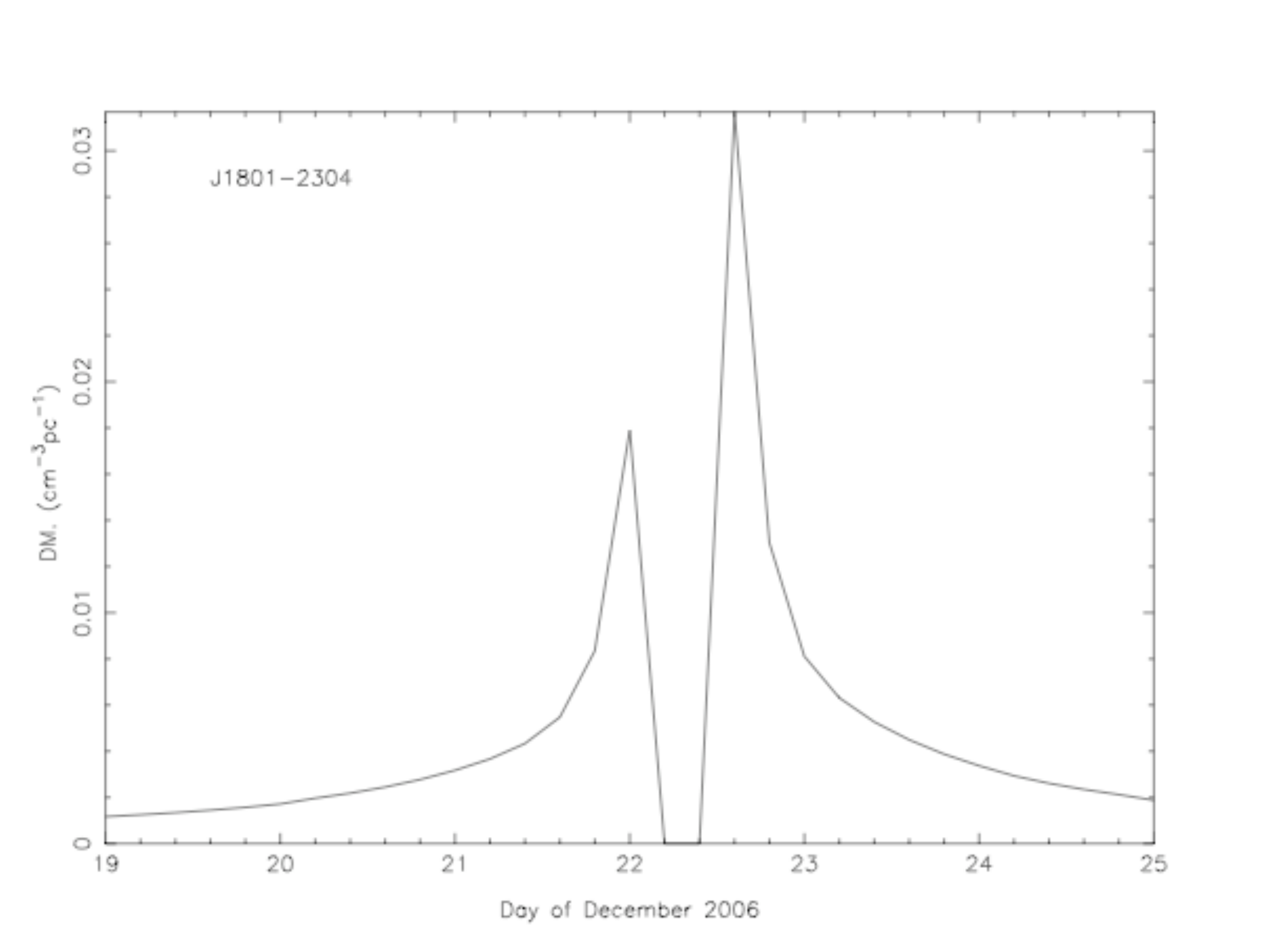}
\caption{The expected DM variations for J1801$-$2304. } 
\label{fig:PSRDM}
\end{figure}

\subsubsection{Rotation Measure}

The modeling has revealed that the expected level of position angle
rotation is incredibly variable, both as a function of observing day and
ecliptic latitude. This is due to the geometry of the coronal magnetic field.
If the line--of--sight to the pulsar projects through a symmetric coronal
magnetic structure then the resultant RM will be negligible. However if the
solar corona is asymmetric the RM, and thus the rotation in PA, can be
considerable. Figure~\ref{fig:PSRFARA} shows this clearly; there is essentially
no rotation in the PA except for one day within the two week period simulated
in the Figure. The results are not unprecedented, Mancuso and Spangler (2000) indicate
variable measured RM (from 0 to 11 rad~m$^{-2}$) at a considerable 
distance from the Sun and Ingleby et al. (2007)\nocite{isw07} detect an RM change of 
61 rad~m$^{-2}$ at a distance greater than 5 solar radii.

J1757$-$2421 also passes close to the solar limb, however it does not 
display significant excess Faraday rotation. The explanation is provided 
by Figure \ref{fig:PSRFARA2} which indicates that the expected Faraday 
rotation is very low. This is due to the line of sight traversing a 
generally symmetric field configuration, which can be discerned from 
Figure \ref{fig:source}. 

\subsubsection{Dispersion Measure}

The essential difference between this experiment and that performed by Mancuso
and Spangler~(2000), is that we are observing simple polarised point sources, with a single dish 
and that we can independently measure the electron component in the coronal
plasma. Pulsar observations typically consider the solar wind as a source of an
additional dispersive delay to that imposed upon observations by the
Interstellar Medium. The analysis of pulse times--of--arrival typically attempt
to remove the additional delay imposed by the solar wind by assuming a very
simple model of the solar corona (see Lommen et al. 2006 \nocite{lkn+06} for a discussion). This has long been considered a gross
simplification by the pulsar community and work is currently underway to
greatly improve upon the current model (Edwards et al. 2006) \nocite{ehm06} . Inadequacies in the current model have
been apparent as periodic DM variations in timing observations of some pulsars
for example Splaver et al. (2005)\nocite{sns+05} present a periodic DM variation of $2 \times 10^{-4}$
pc~cm$^{-3}$ for the pulsar J1713+0747.
Our predictions indicate a level of DM variation at 10 times this level (few
$\times 10^{-3}$~pc~cm$^{-3}$) for some objects, as presented in
Figure~\ref{fig:PSRDM}. This will be challenging to measure for some of the
pulsars in Table~1 as the arrival time precision is not high.

Changes in DM of order 10$^{-3}$ pc~cm$^{-3}$ require a timing precision of
several $\mu$~s at both 10~cm and 50~cm simultaneously. Timing arrival
precision is given, as a rough rule of thumb, by the width of the pulse divided
by the signal--to--noise ratio of the observation. As most of the target list
have widths of order 10~ms, high signal--to--noise ratios will be required to
obtain the required measurement precision. However some of the target pulsars,
notably the millisecond pulsars PSR~J1730--2304 and PSR~J1756--2251, have a
pulse width of order 1~ms and such should allow sufficient measurement
precision.

We were unable to detect any variation in dispersion measure for J1801$-$2304 due to a lack of such precision in time of arrival. Figure~\ref{fig:PSRDM} indicates the expected level of DM variation and as can be seen from Table~\ref{res1801} our DM measurement accuracy falls a long way short.

\subsection{Magnetic Field}

The dispersion measure precision presented in Table~\ref{res1801}
is 0.5~cm$^{-3}$pc, any DM contribution due to the plasma of the solar 
corona must be less than this. We can therefore use this limit to indicate a 
minimum magnetic field (using Equation \ref{B}) of
$>$~393$\mu$G or 3.93nT. 
Under the assumption that the coronal plasma model outlined
in  \S~\ref{electron_density} and shown in Figure~\ref{fig:PSRDM} 
is a reasonable representation of the coronal electron density, we 
can further constrain the integrated magnetic field. Assuming from 
Figure~\ref{fig:PSRDM} the DM contribution on the December 23rd 
is approximately 0.01~cm$^{-3}$pc for the line of sight to J1801$-$2304 
our RM measurement constrains the net integrated line of sight 
magnetic field to be approximately 20mG or 0.2$\mu$T at an 
elongation of 0.96 degrees from the centre of the 
disk (2.7 solar radii from the solar limb). 

\section{Conclusion}
We have presented details of an experiment with the potential to 
independently probe both the electron and magnetic field properties 
within the solar corona. We have proven the practicality of this 
experiment by presenting rotation measure variations detected as 
J1801$-$2304 passed close to the limb of the Sun. These RM measurements 
indicate a net magnetic field of 0.2$\mu$T at 2.7 solar radii
for the solar limb.

This experiment, if conducted in a more complete manner has 
the potential to independently map the coronal plasma 
density and magnetic field along multiple lines of sight. 
As such it is a useful contribution to the current experimental 
probes of coronal properties.
 
\bibliographystyle{aa}
\bibliography{journals_apj,psrrefs,modrefs,crossrefs,thispaper}
\end{article}
\end{document}